\documentclass[symmetry,article,accept,pdftex,oneauthor]{Definitions/mdpi} 

\firstpage{1} 
\pubvolume{0}
\issuenum{0}
\articlenumber{0}
\pubyear{2024}
\copyrightyear{2024}
\externaleditor{Academic Editor: ...}
\datereceived{...} 
\dateaccepted{...} 
\datepublished{...}

\hreflink{https://\linebreak 
	doi.org/...} 
\doinum{...}

\usepackage{bm}
\usepackage{amsmath,amssymb}
\usepackage{wasysym}
\usepackage{pifont,bbm}
\usepackage{stmaryrd}
\newcommand{\xxx}{y_1}
\newcommand{\yyy}{y_2} 
\newcommand{\F}{F}
\newcommand{\B}{G}

\newcommand{\apj}{Astroph. J.}

\newcommand{\jcp}{J. Chem. Phys.}
\newcommand{\be}{\begin{equation}}
\newcommand{\ee}{\end{equation}}

\makeatletter
\renewcommand{\maketag@@@}[1]{\hbox{\m@th\normalsize\normalfont#1}}
\makeatother

\Title{The Schwarzian Approach in Sturm-Liouville Problems}

\TitleCitation{The Schwarzian Approach in Sturm-Liouville Problems}

\Author{Nektarios Vlahakis \orcidA{}}

\AuthorNames{Nektarios Vlahakis}

\AuthorCitation{Vlahakis, N.}

\address[1]{%
Department of Physics, National and Kapodistrian University of Athens, University Campus,\linebreak Zografos GR-157 84 Athens, Greece; vlahakis@phys.uoa.gr
} 

\abstract{ 
A novel method for finding the eigenvalues of a Sturm-Liouville problem is developed. Following the minimalist approach the problem is transformed to a single first-order differential equation with appropriate boundary conditions. Although the resulting equation is nonlinear, its form allows to find the general solution by adding a second part to a particular solution. This splitting of the general solution in two parts involves the Schwarzian derivative, hence the name of the approach. The eigenvalues that correspond to acceptable solutions asymptotically can be found by requiring the second part to correct the diverging behavior of the particular solution. The method can be applied to many different areas of physics, such as the Schr\"odinger equation in quantum mechanics and stability problems in fluid dynamics. Examples are presented. 
}


\begin{document}

\section{Introduction}\label{introduction}

Sturm-Liouville theory is of fundamental importance in many areas of physics and mathematics with important applications in a variety of physical phenomena, in quantum mechanics and in stability problems in fluid mechanics, to name but two. 
Simple examples have become standard knowledge of a physicist, mathematician, or engineer, taught at undergraduate level, while more complicated cases are subject of current research. Standard methods to find solutions have been developed, e.g. \cite{amreinbook,kravchenko2020direct}, but there is always room for improvements and better efficiency in finding the eigenvalues and the corresponding eigenfunctions. 
 
A Sturm-Liouville problem corresponds to a second-order differential equation of the form
\be (pf')'+qf=0
\,,
\label{eqoriginalsl}
\ee
where $p$ and $q$ are functions of the independent variable $x$ (a prime denotes derivative with respect to $x$), together with appropriate boundary conditions that the unknown function $f$ must obey. These boundary conditions involve only the ratios $f'/f$ at the extreme values of $x$ enclosing the region of interest on the $x$ axis. (Depending on the problem this could be the whole axis $-\infty<x<\infty$, a semi-infinite interval $x_{\rm min}<x<\infty$ or $-\infty<x<x_{\rm max}$, or a finite portion $x_{\rm min}<x<x_{\rm max}$.)
Note that the classical Sturm-Liouville problem is written as
$(pf')'+qf=-\lambda w f$, where $w$ is a function of $x$ and $\lambda$ is the eigenvalue. However we can include the right-hand side inside the last term of the left-hand side and write the equation as in (\ref{eqoriginalsl}). Our purpose anyway is to study the more general Sturm-Liouville problem in which the eigenvalue appears in a nonlinear way inside both $q$ and $p$. This is the form arising from the linearization of equations describing stability problems in hydrodynamics and ideal magnetohydrodynamics, see e.g. \cite{Goedbook2,2023Univ....9..386V}. 
Motivated by the need to find an efficient way to solve these problems, trying to identify which factors determine their dispersion relation and exploring ways to express the boundary conditions, we arrive at the novel approach presented in this paper, that applies to all kinds of Sturm-Liouville problems.

In Section \ref{secminimalist} we extend the minimalist approach developed in Ref. \cite{2024Univ...10..183V} to any Sturm-Liouville problem. This provides the more economic way to solve the problem by transforming Equation (\ref{eqoriginalsl}) to a single first-order equation. 
In Section \ref{secschwarzian} we present the new method to split the solution in two parts and use them to conveniently express the boundary conditions corresponding to non-diverging solutions asymptotically. We dub the approach ``Schwarzian'' since the Schwarzian derivative and the fact that it remains unchanged under M\"obius transformations are key ingredients of the developed method. 
In Section \ref{secschwarzian} we apply the method to examples from quantum mechanics.  
In Section \ref{secstability} we discuss the linear stability problems and present their Sturm-Liouville form. An example for the stability of an astrophysical jet is presented in Section \ref{secexamplestability}, using also the new method of solving the complex equation resulting from the boundary conditions developed in Ref. \cite{2024Univ...10..183V}. 
Conclusions follow in Section \ref{secconclusions}.

\section{The Minimalist Approach}\label{secminimalist} 

Since the Equation (\ref{eqoriginalsl}) is linear in $f$, and the boundary conditions involve only the values of the ratios $f'/f$ at the extreme values of $x$, we can reformulate the problem, following the minimalist approach of \cite{2024Univ...10..183V}, using as unknown the ratio of $f$ with its derivative, or equivalently the function 
\be
\F=\dfrac{pf'}{f}
\,. \label{eqdefF}
\ee
It is straightforward to show that Equation (\ref{eqoriginalsl}) is transformed to a first-order differential equation for $\F$
\be
\F'=-\dfrac{\F^2}{p}-q
\,. \label{eqforF}
\ee 
The formulation significantly simplifies the problem of finding the eigenvalues since it transforms the 2nd order original Sturm-Liouville equation to a 1st order. The new equation is non-linear, but this does not complicate things if one uses a shooting method to satisfy the boundary conditions; it only makes the procedure simpler by reducing the number of first-order differential equations to half (the original second-order equation can be thought as two first order).

In some cases, especially when we integrate problems in real space, possible oscillations in $f$ lead to $\F$ varying from infinity to infinity and back. One way to treat these cases and pass smoothly infinities at points where $f=0$ is to substitute $\F(x)=\cot\dfrac{\Phi(x)}{2}$.
More generally we can write 
\be \F(x)=\F_1(x)+\F_2(x)\cot\dfrac{\Phi(x) }{2}
\ee 
with given $\F_1(x)$ and $\F_2(x)$ of our choice. 
In this way the zeros of $f$ correspond simply to $\Phi$ being an even multiple of $\pi$ and the solutions can be found without any numerical difficulty by integrating the differential equation 
\be \Phi'=\dfrac{2\F_1\F_2+p\F_2'}{p\F_2} \sin\Phi-\dfrac{pq+p\F_1'+\F_1^2-\F_2^2}{p\F_2} \cos\Phi
+\dfrac{pq+p\F_1'+\F_1^2+\F_2^2}{p\F_2}
\,. 
\label{eqforPhiAtanB}
\ee

The boundary conditions for $F$ are translated to boundary conditions for $\Phi$, and they are satisfied for the eigenvalues of the problem.
Having found an eigenvalue one could return to the original equation, or simply to Equation~(\ref{eqdefF}) which gives $\dfrac{f'}{f}=\dfrac{\F}{p}$, and by integration find $f$. (Although $F$ and $\Phi $ are uniquely defined for a particular eigenvalue, there is a free multiplication constant in the eigenfunction $f$, which is related to the normalization.) 

\section{The Schwarzian Approach}\label{secschwarzian} 
\subsection{The Schwarzian $g$ Approach} 

We can once more reformulate the problem and write its general solution using the Schwarzian derivative. As it will become clear later, this is particularly useful in cases where the problem extends to asymptotic regions, as the new approach is linked to conveniently satisfying the boundary conditions there, by automatically choosing the non-diverging solution.

It is straightforward to show that Equation (\ref{eqoriginalsl}) can be transformed to a ``variable frequency oscillator'' 
\be (f\sqrt{p})''+\kappa^2(f\sqrt{p})=0
\label{eqforb}
\ee 
with \be \kappa^2=\dfrac{q}{p}-\left(\dfrac{p'}{2p}\right)^2-\left(\dfrac{p'}{2p}\right)' \,. \ee
As explained in Appendix \ref{appendixschwarzian} we can write its general solution as 
\be f=\dfrac{C_1g+C_2}{\sqrt{pg'}} 
\,, \label{eqsolforb} \ee 
where $g$ is a particular solution of the Schwarz equation
\be \{g,x\}\equiv \frac{g'''}{g'} - \frac{3}{2}\left(\frac{g''}{g'}\right)^2 =2\kappa^2
\,, \label{eqgappr}\ee 
involving the Schwarzian derivative $\{g,x\}$. 

We can transform the latter to a system of first order differential equations 
by writing the expression for $\F=\dfrac{pf'}{f}=-\dfrac{pg''}{2g'}-\dfrac{p'}{2}+\dfrac{pg'}{g+C_2/C_1}$ 
(which is the general solution of Equation \ref{eqforF}, since Equation \ref{eqsolforb} is the general solution of Equation \ref{eqoriginalsl})
as
\begin{eqnarray}
	\F=\F_p+\dfrac{e^{-2\Lambda}}{g+C_2/C_1} \,,  
	\label{eqsgScwarzianF}
\end{eqnarray}
where $\F_p=-\dfrac{pg''}{2g'}-\dfrac{p'}{2}$ and $e^{-2\Lambda}=p g' $.
The last two equations together with Equation~(\ref{eqgappr}) give the system
\begin{eqnarray} 
\F_p'=-\dfrac{\F_p^2}{p}-q \,, \quad
\Lambda'=\dfrac{\F_p}{p} \,,\quad 
g'=\dfrac{e^{-2\Lambda}}{p} \,.
\label{eqsgScwarzian}
\end{eqnarray}
The solution of the Sturm-Liouville equation is
\be f=\dfrac{C_1g+C_2}{e^{-\Lambda}} 
\,, \label{eqsolforbF2} \ee
we remind though, that to find the eigenvalues only $\F$ is needed.

Interestingly the first of Equations (\ref{eqsgScwarzian}) is the same as the original Equation (\ref{eqforF}), meaning that $F_p$ is a particular solution of that equation, not necessarily satisfying the boundary conditions.
These conditions should be satisfied by the total $F$ given by Equation (\ref{eqsgScwarzianF}), consisting of two parts. Its second part can be found using the second and third of Equations (\ref{eqsgScwarzian}). We emphasize that only a particular solution is needed, the initial values of $\F_p$, $\Lambda$, $g$ at some initial point of integration are completely free. For any choice of these conditions we find $F$ and apply the two boundary conditions at the ends of the region of interest. One of them specifies the free constant $C_2/C_1$ appearing in Equation (\ref{eqsgScwarzianF}) and the other gives the eigenvalues.

\subsection{The Schwarzian $\Phi$ Approach}

An alternative way to solve the problem, particularly useful if there are points where $f=0$ (so $\F$ becomes infinity), but in general with some potential connection to the phase of oscillations of the function $f$, is to replace $g+\dfrac{C_2}{C_1}$ with $\tan\dfrac{\Phi+C}{2}$
(and constant $C$).     
Substituting $g$ in the expressions above and using the chain rule \ref{eqchainrule} with $\{g,\Phi\}=\dfrac{1}{2}$, 
we conclude that the general solution of the ``variable frequency oscillator'' is 
\be f\propto \dfrac{1}{\sqrt{p\Phi'}}\sin\dfrac{\Phi+C}{2}  \,,
\label{eqforbPhi}\ee 
and $\Phi$ a particular solution of 
\be \left\{\tan\dfrac{\Phi+C}{2},x\right\}\equiv \dfrac{\Phi'''}{\Phi'}-\dfrac{3\Phi''^2}{2\Phi'^2}+\dfrac{\Phi'^2}{2} =2\kappa^2
\,. \label{eqPhiappr} \ee 
 
We can transform the latter to a system of first order differential equations 
by writing the expression for $\F=\dfrac{pf'}{f}=-\dfrac{p\Phi''}{2\Phi'}-\dfrac{p'}{2}+\dfrac{p\Phi'}{2}\cot\dfrac{\Phi+C}{2}$ as 
\begin{eqnarray}
	\F=\F_1+\F_2\cot\dfrac{\Phi+C}{2} \,,  
	\label{eqforFPhi}
\end{eqnarray}
where $\F_1=-\dfrac{p\Phi''}{2\Phi'}-\dfrac{p'}{2}$ and $\F_2=\dfrac{p\Phi'}{2}$.
The last two equations together with Equation~(\ref{eqPhiappr}) give the system
\begin{eqnarray} 
	\F_1'= \dfrac{ \F_2^2}{p}-\dfrac{\F_1^2}{p}-q \,, 
	\quad
	\F_2'=-\dfrac{2\F_1\F_2}{p} \,, 
	\quad
	\Phi'=\dfrac{2\F_2}{p}\,.
	\label{eqsPhiScwarzian}
\end{eqnarray}
The solution of the Sturm-Liouville equation is (with $D$ a normalization constant)\endnote{{Some caution is needed when one calculates the square root $ \sqrt{\F_2}$ in Equation (\ref{eqforbPhiF2}). Since the constant $D$ is arbitrary we are free to chose e.g. the principal square root. In cases however in which as $x$ varies the ${\rm Arg} [\F_2]$ crosses the value $\pi$ we should change branch to avoid a false discontinuity in $\Im[\sqrt{\F_2}]$.}}
\be f=\dfrac{D}{\sqrt{\F_2}}\sin\dfrac{\Phi+C}{2}  \,.
\label{eqforbPhiF2} \ee
(We again remind that to find the eigenvalues only $\F$ is needed.) 

The equations of the Schwarzian $\Phi$ approach are of course equivalent to the ones of the $g$ approach, they correspond to the substitutions $g+\dfrac{C_2}{C_1}= \tan\dfrac{\Phi+C}{2}$,
$e^{-2\Lambda}= \dfrac{\F_2}{\cos^2\dfrac{\Phi+C}{2}}$, $\F_p= \F_1-\F_2\tan\dfrac{\Phi+C}{2}$. Note also that $\tan\dfrac{\Phi+C}{2}$ can be seen as a M\"obius transformation of $e^{-i\Phi}$, and the substitutions $g+\dfrac{C_2}{C_1}=e^{-i\Phi}-e^{iC}$, $e^{-2\Lambda}=-2i\F_2e^{-i\Phi}$, $\F_p=\F_1+i\F_2$ lead also to the equations of the Schwarzian $\Phi$ approach.

It is interesting to note the connection of the Schwarzian $\Phi$ approach with the one described at the end of Section~\ref{secminimalist}.
If in that approach we substitute $\Phi\to \Phi+C $, and require the $\Phi'$ as given by Equation~(\ref{eqforPhiAtanB}) to be independent of $\Phi$ (i.e., choose $\F_1$ and $\F_2$ such that the coefficients of $\sin\Phi$ and $\cos\Phi$ are zero), we arrive at the Schwarzian $\Phi$ approach.
Similarly the substitution $\F=\F_p+\dfrac{e^{-2\Lambda}}{g}$ in Equation~(\ref{eqforF}) gives 
$g'=\dfrac{e^{-2\Lambda}}{p}+\left(-2\Lambda' +\dfrac{2\F_p }{p}\right)g+\left(q +\F_p'+\dfrac{\F_p^2}{p}\right)\dfrac{g^2}{e^{-2\Lambda}}$.
The replacement $g\to g+C_2/C_1$, and most importantly the requirement the $g'$ to be independent of $g$, leads to the Schwarzian $g$ approach and Equations~(\ref{eqsgScwarzian}).

\subsection{Non-Diverging Asymptotic Solutions}\label{secNon-Diverging}

Besides offering a way to express the general solution of a Sturm-Liouville problem, the Schwarzian approach can be used to choose the non-diverging asymptotically solution of a physical problem.
In the $g$ approach this is achieved through the minimalist approach (solve the solution for $\F$) and the splitting of $F$ in two parts as shown in Equation~(\ref{eqsgScwarzianF}).

Suppose that we examine the behavior of the solution near $x=+\infty$. The two independent solutions of Equation (\ref{eqoriginalsl}), or equivalently of Equation (\ref{eqforb}), behave as $\dfrac{1}{\sqrt{p_\infty}}e^{\pm i\kappa_\infty x}=\dfrac{1}{\sqrt{p_\infty}}e^{\pm i\Re\kappa_\infty x}e^{\mp \Im\kappa_\infty x}$. One of them is diverging and dominates any linear combination in the general solution. This is the case for the part $\F_p$ of the solution which by itself satisfies the original Equation (\ref{eqforF}), see the first from Equations (\ref{eqsgScwarzian}). 
The diverging solution corresponds to positive real part of $\F_p/p$ and thus exponentially decreasing $e^{-2\Lambda}$ according to the middle from Equations (\ref{eqsgScwarzian}). Although $e^{-2\Lambda}$ vanishes asymptotically, the freedom to choose the additive constant $C_2/C_1$ in Equation (\ref{eqsgScwarzianF}) 
-- essentially the fact that the Schwarzian derivative remains the same under M\"obius transformations -- 
allows to make the second part of the solution important. By choosing $g+C_2/C_1\to 0$ this second part $\dfrac{e^{-2\Lambda}}{g+C_2/C_1}$ becomes $\dfrac{0}{0}$, and by applying L'H\^opital's rule we find that it
equals $\dfrac{-2\Lambda' e^{-2\Lambda}}{g'}=-2\F_p$. The resulting sum follows exactly the non-diverging branch since $\F_p+\dfrac{e^{-2\Lambda}}{g+C_2/C_1}\to -\F_p$.

It is instructive to exactly solve the case $p=1$, $q=$ constant, corresponding to an oscillator with complex constant ``frequency'' $\kappa=\sqrt{q}$. Without loss of generality we can choose the root with positive $\Im\kappa$. 
A particular solution of Equations~(\ref{eqsgScwarzian}) is $\F_p=-\kappa \tan(\kappa x)$, $e^{-2\Lambda}=\dfrac{1}{\cos^2(\kappa x)}$, $g=\dfrac{\tan(\kappa x)}{\kappa}$.
Asymptotically $\displaystyle\lim_{x\to+\infty} g=\dfrac{i}{\kappa}$. Choosing the additive constant such that $\dfrac{i}{\kappa}+\dfrac{C_2}{C_1}=0$ we find
$\F=\F_p+\dfrac{e^{-2\Lambda}}{g-i/\kappa} =i\kappa$.
We see that although the random combination of the independent solutions corresponds to $\F_p$ and the diverging solution $\F_p=\dfrac{pf'}{f}\Rightarrow f=\exp \left(\displaystyle\int \F_p dx\right)\propto \cos(\kappa x)$, the solution corresponding to $\F=i\kappa$ is $\F=\dfrac{pf'}{f}\Rightarrow f=\exp \left(\displaystyle\int \F dx\right)\propto e^{i\kappa x}$, and has the physically acceptable behavior $\displaystyle\lim_{x\to+\infty} f=0$.\endnote{{The general solution of Equations~(\ref{eqsgScwarzian}) is $\F_p=-\kappa \tan(\kappa x+D_1)$, $e^{-2\Lambda}=\dfrac{D_2}{\cos^2(\kappa x+D_1)}$, $g=D_2\dfrac{\tan(\kappa x+D_1)}{\kappa}+D_3$.
Repeating the process we again find $\F=i\kappa$, confirming that only a particular solution is needed, not the general.}}

The above can be applied to other asymptotic regimes, e.g. to $x\to-\infty$. They can also be applied in the asymptotic regimes $\varpi\to\infty$ and $\varpi\to 0$ in cylindrical coordinates.

Similarly to the Schwarzian $g$ approach in the $\Phi$ approach
Equation (\ref{eqforFPhi}) can be written
\be \F=\left(\F_1+i\F_2\right)+\dfrac{2i\F_2}{e^{i(\Phi+C)}-1}
\label{eqsplitiphi}
\ee 
and be seen as a way to split the solution in two parts. The first  
$\F_1+i\F_2$ satisfies by itself the original Equation (\ref{eqforF}) if Equations (\ref{eqsPhiScwarzian}) hold, and corresponds to diverging solution asymptotically. 
The second part $\dfrac{2i\F_2}{e^{i(\Phi+C)}-1}$
can be used to find the non-diverging solution, by choosing the constant $C$ such that $e^{i(\Phi+C)}=1$ asymptotically. 

Had we chose the Schwarzian $\Phi$ approach to solve the oscillator problem (with constant $p=1$, $q=\kappa^2$, and $\Im\kappa>0$), starting the integration from some point with arbitrary conditions, the solution would behave asymptotically as $\Phi\approx De^{2i\kappa x}+D_0$,
$\F_1\approx-i\kappa$, $\F_2\approx i\kappa D e^{2i\kappa x}$.\endnote{{The general solution is 
$\cot\dfrac{\Phi+D_3}{2}=D_2 \cot(\kappa x+D_1) $.}}
The resulting $F=-i\kappa+i\kappa D e^{2i\kappa x} \cot \dfrac{De^{2i\kappa x}+D_0+C}{2}$, for $D_0+C=0$, would have given the physically acceptable solution $F=i\kappa$.

\subsection{The Quantization Condition}

Suppose we follow the Schwarzian $\Phi$ approach (the steps are trivially similar if we follow the Schwarzian $g$ approach). 
Solving Equation (\ref{eqPhiappr}) with arbitrary initial conditions we find sort of a background for the solution for $\F$. (We remind that to find a particular solution is enough, there is no need to find the general solution.) 
This is because the expression for $\F$ given by Equation~(\ref{eqforFPhi}) depends on the free constant $C$, so by appropriately choosing this constant we can satisfy one of the boundary conditions of the problem.  

There is always a second boundary condition to satisfy, and this can be done with the shooting method, i.e., change the eigenvalue until the second condition is also satisfied (for each eigenvalue the constant $C$ is found from the first boundary condition). 

It is of particular importance to discuss the case where one (or both) boundary is at infinity.
As already discussed the second order differential Equation (\ref{eqoriginalsl}) has asymptotically two solutions, one diverging and unphysical, and another finite and physically acceptable. Their superposition is diverging and for this reason it is nontrivial to find the acceptable solution. 
The Schwarzian approach solves the problem by splitting the solution $F$ in two parts, see Equation (\ref{eqforFPhi}). The part $\F_1$ is dominated by the diverging behavior asymptotically. The inclusion of the second part of the solution $\F_2\cot\dfrac{\Phi+C}{2} $ is the way to correct things and have the sum following the physically acceptable, finite solution. Although $\F_2$ vanishes asymptotically 
(if $\F_1=pf_1'/f_1$ we get $\F_2\propto 1/f_1^2$ from the middle form Equations~\ref{eqsPhiScwarzian}, and thus if $f_1$ diverges $\F_2$ vanishes), this is achieved by requiring infinite $\cot\dfrac{\Phi+C}{2} $, i.e., have a ``quantization'' condition of $\Phi +C$ being equal to an even multiple of $\pi$ asymptotically. 
\\ From a different point of view related to causality, if we want the two boundary conditions to ``communicate'', the term in Equation (\ref{eqforFPhi}) that includes $C$, which contains the information of the first boundary condition, should remain present at the location of the second boundary condition. If $\Phi'$ vanishes this is possible only if $\cot\dfrac{\Phi+C}{2}$ diverges.
\\ Another way to reach the same result is by observing Equation~(\ref{eqforbPhi}) or (\ref{eqforbPhiF2}). If $\Phi'$ vanishes asymptotically, the only way to have finite $f$ is by requiring $\sin\dfrac{\Phi+C}{2}$ to be zero as well. 
\\ Lastly, looking Equation (\ref{eqforb}) the second part remains important and corrects the diverging behavior of the first if asymptotically $e^{i(\Phi+C)}=1$.
\\  
For all these reasons the related boundary condition is:
\be 
\mbox{if } \Phi_{\rm BC}'=0 \mbox{ then } \Phi_{\rm BC}+C=2n\pi \,, n\in \mathbb{Z}
\,. \ee

If the two boundaries are at $\pm \infty$ then the quantization condition is that $\Phi+C$ equal to an even multiple of $\pi$ at both ends, so the difference is also an even multiple of $\pi$. Instead of both conditions the eigenvalues can be found using the following single quantization condition:
\be 
\mbox{if } \Phi|_{1,2}'=0 \mbox{ then } \Phi|_{2}-\Phi|_{1}=2n\pi \,, n\in \mathbb{Z}
\,. \ee

The corresponding quantization conditions in the Schwarzian $g$ approach are:
\be 
\mbox{if } g_{\rm BC}'=0 \mbox{ then } g_{\rm BC}+C_2/C_1=0
\,, \ee
\be 
\mbox{if } g|_{1,2}'=0 \mbox{ then } g|_{2}-g|_{1}=0
\,. \ee

The quantization conditions, and the Schwarzian approaches in general, offer a convenient way to solve the problem exactly in cases where the regime of interest extend to asymptotic regions where all functions approach constant values. This applies to $x=\pm\infty$ in cartesian geometry, but also to $\varpi\to 0,\infty$ in cylindrical, and to $r\to 0,\infty$ in spherical (with $\varpi$ and $r$ the cylindrical and spherical radius, respectively). 
When the regime of interest does not include asymptotic regions it is sufficient to follow the minimalist approach of Section \ref{secminimalist}. Of course Schwarzian approaches still can be used, but the boundary conditions are $\F|_{1,2}=\F_{\rm BC }|_{1,2}$ at the two extreme values of the independent variable ($x$ or $\varpi$ or $r$), and not the quantization conditions.

\section{Examples in Quantum Mechanics}\label{secexqm}

\subsection{Quantum Morse Potential}\label{secquantumMorse}

The Schr\"odinger equation $-\dfrac{\hbar^2}{2m}\Psi''+(V-E)\Psi=0$ is a Sturm-Liouville problem, and with proper normalization the Morse potential corresponds to the choice $p=1$, $q=\varepsilon-\lambda^2\left(1-e^{-x}\right)^2$
(the origin has been transfered at the minimum of the potential).
The eigenvalues can be easily found following the Scharzian $\Phi$ approach, i.e., integrating Equations (\ref{eqsPhiScwarzian}) starting from $x=0$, moving toward positive and negative $x$, and requiring
the difference $\dfrac{\Phi|_{+\infty}-\Phi|_{-\infty}}{2\pi}$ to be an integer.  

\begin{figure}[H] 
	\includegraphics[width=.4\textwidth]{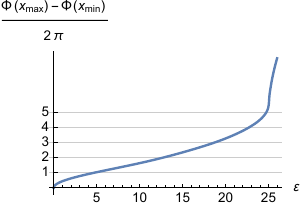}
	\hfill
	\includegraphics[width=.5\textwidth]{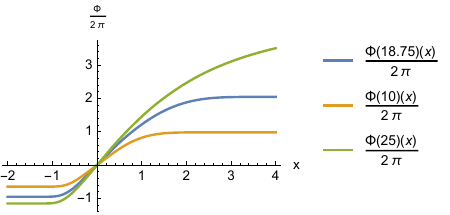}
	\\
	\includegraphics[width=.4\textwidth]{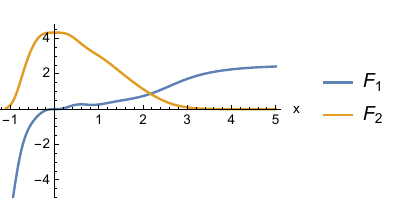}
	\hfill
	\includegraphics[width=.55\textwidth]{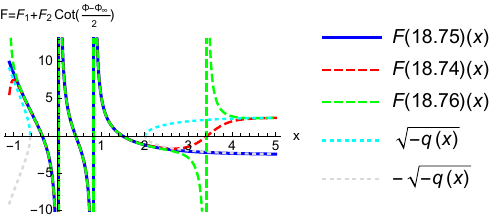}
	\caption{Solution for the Morse potential with $\lambda=5$,
		for initial conditions $\Phi(0)=0$, $\F_1(0)=0$, $\F_2(0)=\sqrt{q(0)}$.
		Top left panel: The value of  $\dfrac{\Phi|_{+\infty}-\Phi|_{-\infty}}{2\pi}$ as function of $\varepsilon$ (integer values correspond to the accepted eigenvalues).
		Top right panel: Typical behavior of $\Phi(x)$ (shown for various values of the eigenvalue $\varepsilon$).
		Bottom left panel: The functions $\F_1$ and $\F_2$ for the third eigenvalue $\varepsilon=18.75 $ (corresponding to $n=2$).
		Bottom right panel: The function $\F$ for the eigenvalue $\varepsilon=18.75$ and two neighboring values of $\varepsilon$. 
		The function $F$ approaches asymptotically one of the $\pm\sqrt{-q}$, that are also shown. 
		At large $x$ the blue curve follows the acceptable branch with $f'/f \approx -\sqrt{-q(x)}$, while the red and green follow the unphysical branch with $f'/f\approx \sqrt{-q(x)}$.
		\label{figmorse}
	}
\end{figure}

Figure \ref{figmorse} shows the numerical results for $\lambda=5$. The integration in the region $-7<x<15$ is sufficient to give the eigenvalues (we need to resolve the region in which $\F_2$ remains practically nonzero and not the whole $x$ axis). 
The results agree with the known expression for the eigenvalues $\varepsilon= \lambda^2-(\lambda-n-1/2)^2 $ with $n=0,1,2,\dots$ and positive $\lambda-n-1/2$ (corresponding to bound states with energy less than $V_\infty$), see e.g. \cite{1988JChPh..88.4535D}.
The quantum number $n$ corresponds to the integer $\dfrac{\Phi|_{+\infty}-\Phi|_{-\infty}}{2\pi}-1$.

The bottom right panel of the Figure \ref{figmorse} shows the function $\F$.
Asymptotically, according to Equation~(\ref{eqforF}), we expect the function $\F$ to approach either $+\sqrt{-pq}$ or $-\sqrt{-pq}$. As $x\to\infty$ these correspond to positive/negative $f'/f$, i.e., an exponentially increasing/decreasing $f$, respectively. Any superposition of the two solutions of the second order differential Equation (\ref{eqoriginalsl}) is dominated by the exponentially increasing part, and is unacceptable as physical solution of the problem. The acceptable solution corresponds to cases where the exponentially increasing part is absent, and this is what happens for the eigenvalues. Indeed as shown in the Figure, the blue curve which corresponds to the eigenvalue $n=2$ follows the acceptable branch $\F\to-\sqrt{-q}$ while the red and green curves which correspond to slightly smaller and larger value of $\varepsilon$, respectively, 
follow the unphysical branch $\F\to +\sqrt{-q}$. 
Similarly we understand the behavior of $\F$ in the asymptotic regime $x\to-\infty$. 

The bottom left panel of the Figure \ref{figmorse} shows the functions $\F_1$ and $\F_2$.
To former corresponds to a superposition that includes the unphysical branch and behaves asymptotically as $\approx +\dfrac{x}{|x|}\sqrt{-pq}$, while the latter approaches zero at $x\to\pm\infty$. Nevertheless, its contribution is crucial and results in $\F=\F_1+\F_2\cot\dfrac{\Phi+C}{2}$ following the physical acceptable branch if $\varepsilon$ is an eigenvalue.

\begin{figure}[H]
	\includegraphics[width=.4\textwidth]{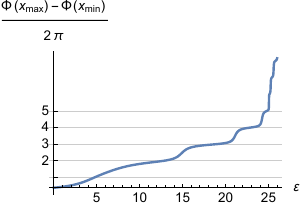}
	\hfill
	\includegraphics[width=.5\textwidth]{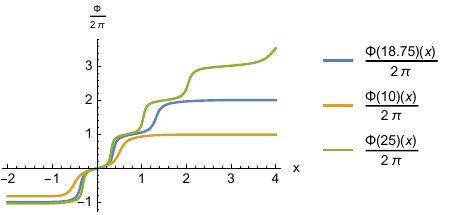}
	\\
	\includegraphics[width=.4\textwidth]{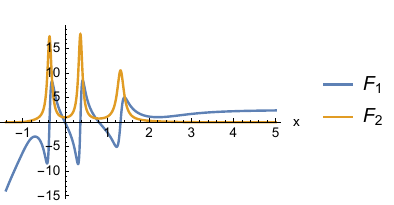}
	\hfill
	\includegraphics[width=.55\textwidth]{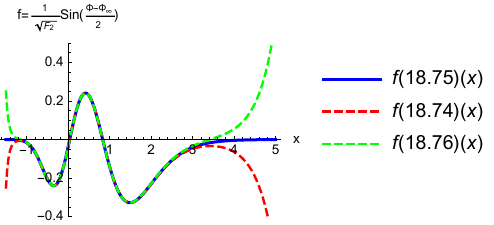}
	\caption{Same as Figure \ref{figmorse}, but for a boundary condition $\F_2(0)=1$. The functions $\Phi(x)$, $\F_1(x)$, $\F_2(x)$ are affected by the choices of the conditions at $x=0$, but the difference $\Phi|_{+\infty}-\Phi|_{-\infty}$ and the function $\F(x)$ are the same for the same eigenvalue $\varepsilon$. The eigenfunction $f(x)$
		(with arbitrary normalization) for the eigenvalue $\varepsilon=18.75$ and two neighboring values of $\varepsilon$ is shown in the bottom right panel.
		\label{figmorsediffslope}
	}
\end{figure} 

The eigenvalues do not depend on the initial conditions at $x=0$. Our choice in Figure~\ref{figmorse} was $\F_1(0)=0$, $\F_2(0)=\sqrt{q(0)}$ such as $\F_1'(0)=0$ and thus the value of $\Phi'$ to remain as close to a constant as possible, since $\Phi''(0)=0$ and $\Phi'''(0)=0$. 
In the general case the choice $\Phi''=0$, $\Phi'''=0$ at the initial point is possible and corresponds to initial values $\F_1=-\dfrac{p'}{2}$, $\F_2=p\kappa $.
This is not necessary but it is a convenient choice avoiding as much as possible oscillations in the functions $\F_{1,2}$. In the shown case the oscillations in $\F$ are solely due to the $\tan$ function.
On the contrary, Figure \ref{figmorsediffslope} shows the numerical results for different initial conditions and we see that $\F_{1,2}$ oscillate. The end result in the function $\F$ and the eigenvalues are the same as expected, independent on the initial conditions of the integration. Since $F$ is the same as in Figure \ref{figmorse} we show in the bottom right panel of Figure \ref{figmorsediffslope} the eigenfunction $f$ (with arbitrary normalization). 
The blue curve corresponds to the eigenvalue while the red and green curves to two neighboring values of $\varepsilon$ (clearly $f$ diverges at $x\to\pm \infty$ if $\varepsilon$ is not an eigenvalue).   

\begin{figure}[H]
	\includegraphics[width=.4\textwidth]{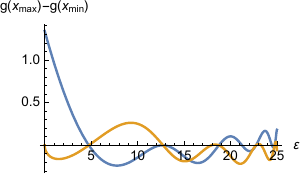}
	\hfill
	\includegraphics[width=.4\textwidth]{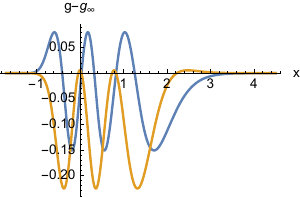}
	\smallskip \\
	\includegraphics[width=.4\textwidth]{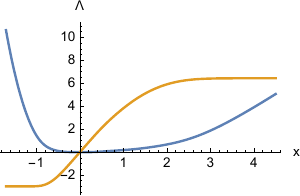}
	\hfill
	\includegraphics[width=.4\textwidth]{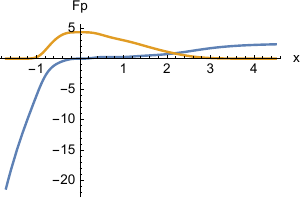}
	\smallskip \\
	\includegraphics[width=.4\textwidth]{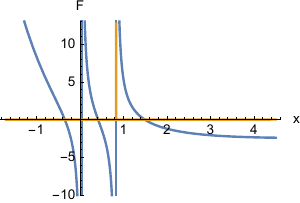}
	\hfill
	\includegraphics[width=.41\textwidth]{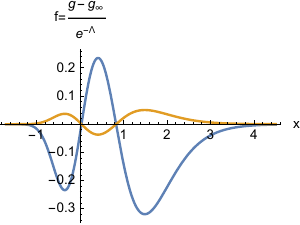}
	\caption{Solution for the Morse potential using the Schwarzian $g$ approach.
		In all panels the real part of the functions is shown with blue lines and the imaginary part with orange. 
		The eigenvalues for which $g|_{+\infty}-g|_{-\infty}=0$
		(practically $g(x_{\rm max})-g(x_{\rm min})=0$ with $x_{\rm min}=-7$, $x_{\rm max}=15$) can be seen in the top left panel.
		The other panels correspond to the solution for the third eigenvalue $\varepsilon=18.75$. 
		\label{figmorseoscilg}}
\end{figure}

The same results can be found using the Schwarzian $g$ approach.
	The problem of the oscillations that make $F$ infinity at points where $f=0$ can be circumvented by working in the complex domain. Figure \ref{figmorseoscilg} shows the results of the integration of Equations (\ref{eqsgScwarzian}) in the interval $-7<x<15$ with initial conditions $g(0)=0$, $\Lambda(0)=0$ and $F_p(0)=i\sqrt{q(0)}$. The resulting $F$, given by Equation~(\ref{eqsgScwarzianF}) with $C_2/C_1=-g|_{-\infty}$, is the same as the one found using the $\Phi$ approach.
	The eigenfunction $f$ (given by Equation \ref{eqsolforbF2}) is also the same, apart from a multiplication complex constant.

\subsection{Quantum Harmonic Oscillator}\label{secquantumharmonic}

\begin{figure}[H]
	\includegraphics[width=.4\textwidth]{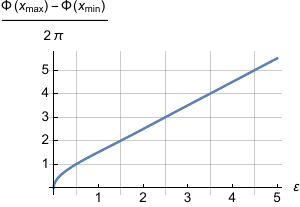}
	\hfill
	\includegraphics[width=.55\textwidth]{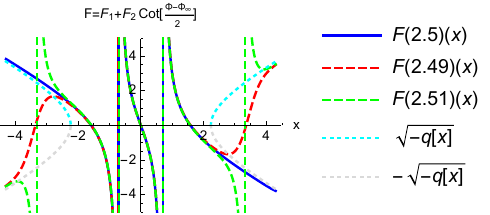}
	\caption{Solution for the quantum harmonic oscillator.
		Left panel: The value of  $\dfrac{\Phi|_{+\infty}-\Phi|_{-\infty}}{2\pi}$ as function of the eigenvalue.
		Right panel: The function $\F$ for the eignevalue $\varepsilon=2.5$ (in blue) and two neighboring values (in red and green). The function $F$ approaches asymptotically one of the $\pm\sqrt{-q}$, that are also shown.
		The blue curve follows the acceptable branch with $f'/f \approx -x$ at large $|x|$, while the red and green follow the unphysical branch with $f'/f\approx x$. 
	\label{figharmonicoscil}}
\end{figure}

The Schr\"odinger equation for a harmonic potential corresponds to the choice $p=1$, $q=2\varepsilon-x^2$ (with proper normalization).
Following the Schwarzian $\Phi$ approach, integrating Equations (\ref{eqsPhiScwarzian}) starting from $x=0$, and requiring the difference $\dfrac{\Phi|_{+\infty}-\Phi|_{-\infty}}{2\pi}$ to be an integer, we find the expected eigenvalues $\varepsilon=n+1/2$ as shown in Figure~\ref{figharmonicoscil}.

\subsection{Paine Problem}

\begin{figure}[H]
	\includegraphics[width=.4\textwidth]{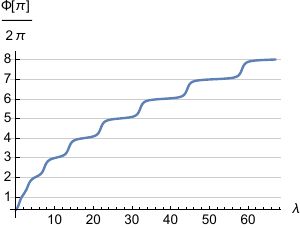}
	\hfill
	\includegraphics[width=.5\textwidth]{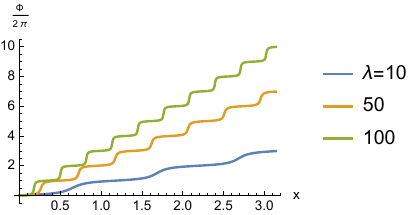}
	\caption{Left panel: The value of  $\dfrac{\Phi(\pi)}{2\pi}$ as function of the eigenvalue.
		Right panel: Typical behavior of $\Phi(x)$ (shown for various values of the eigenvalue $\lambda$).} 
	\label{figPaine}
\end{figure}

This is a test spectral problem corresponding to $p=1$, $q=\lambda-\dfrac{1}{(x+0.1)^2}$ with boundary conditions $f(0)=f(\pi)=0$, see Ref. \cite{kravchenko2020direct}.
Since the region of interest does not contain asymptotic regimes,
it is sufficient to use the minimalist approach (and not the Schwarzian). 
To avoid infinities at possible points where $f=0$, especially since we work in real space, it is better to use the method described at the end of Section~\ref{secminimalist}.
The boundary conditions are $\F(0)=\F(\pi)=\infty$.
Adopting the simplest possible choice $\F_1(x)=0$, $\F_2(x)=1$,
the boundary conditions are $\cot\dfrac{\Phi(0)}{2}=\cot\dfrac{\Phi(\pi)}{2}=\infty$. Equation~(\ref{eqforPhiAtanB}) can be integrated in the interval $x\in [0,\pi]$ starting with $\Phi(0)=0$.
The second condition corresponds to $\Phi(\pi)=2n\pi$ with integer $n$, a condition that gives the eigenvalues.
The numerical results are shown in Figure~\ref{figPaine}.
The eigenvalues are 
$\lambda\approx 1.51987$, $4.94331$, $ 10.2847$, $17.5599$, $
26.7828$, $37.9643$, $
51.1131$, $66.2361$, $ 
83.3385$, $102.424$, $ 
123.497$, $ 146.558$, $ 171.611$, $198.657$,~\dots

\section{Stability Problems}\label{secstability} 
The stability problems in fluid or plasma dynamics can be formulated using two functions, $\xxx$ related to the Lagrangian displacement of fluid elements, and $\yyy$ related to the perturbation of the total pressure. The corresponding system of differential equations can be found, e.g., in Equation (27) of Ref. \cite{2023Univ....9..386V}
\begin{eqnarray} 
	\dfrac{d}{d\varpi} \left( \begin{array}{c}
		\xxx \\ \yyy
	\end{array} \right) + \dfrac{1}{{\cal D}} \left( \begin{array}{cc}
		{\cal F}_{11} & {\cal F}_{12} \\ {\cal F}_{21} & {\cal F}_{22}
	\end{array} \right) \left( \begin{array}{c}
		\xxx \\ \yyy
	\end{array} \right)
	=0\,.
	\label{systemodes}
\end{eqnarray}
This applies to non-relativistic and relativistic ideal magnetohydrodynamic cases in cylindrical geometry in which the 
unperturbed state depends only on the cylindrical radius $\varpi$; 
a similar system of equations exists in other geometries as well, e.g. in cartesian geometry with the unperturbed state depending on $x$ (actually the Equations of \cite{2023Univ....9..386V} can be directly simplified to that case if we ignore terms that depend on the curvature of the coordinate system). 

Instead of the system (\ref{systemodes}) of the two first-order differential equations for $\xxx$ and $\yyy$, it is equivalent to work with one second order differential equation for either $\xxx$ or $\yyy$, i.e., use either Equations (32), or (33) of \cite{2023Univ....9..386V}
\begin{adjustwidth}{-\extralength}{0cm}\begin{eqnarray}
		\xxx''
		+\left[\dfrac{{\cal F}_{11}+{\cal F}_{22}}{{\cal D}}
		+\dfrac{{\cal F}_{12}}{{\cal D}} \left(\dfrac{{\cal D}}{{\cal F}_{12}}\right)'\right]\xxx'
		+\left[\dfrac{{\cal F}_{11} {\cal F}_{22}-{\cal F}_{12} {\cal F}_{21}}{{\cal D}^2}
		+\dfrac{{\cal F}_{12}}{{\cal D}} \left(\dfrac{{\cal F}_{11}}{{\cal F}_{12}}\right)'\right]\xxx =0
		\,, \quad 
		\yyy=-\dfrac{{\cal D}\xxx'+{\cal F}_{11} \xxx}{{\cal F}_{12}}
		\,,
		\label{odefory1}
\end{eqnarray}\end{adjustwidth}
\begin{adjustwidth}{-\extralength}{0cm}\begin{eqnarray}
		\yyy''
		+\left[\dfrac{{\cal F}_{11}+{\cal F}_{22}}{{\cal D}}
		+\dfrac{{\cal F}_{21}}{{\cal D}} \left(\dfrac{{\cal D}}{{\cal F}_{21}}\right)'\right]\yyy'
		+\left[\dfrac{{\cal F}_{11} {\cal F}_{22}-{\cal F}_{12} {\cal F}_{21}}{{\cal D}^2}
		+\dfrac{{\cal F}_{21}}{{\cal D}} \left(\dfrac{{\cal F}_{22}}{{\cal F}_{21}}\right)'\right]\yyy =0
		\,, \quad 
		\xxx=-\dfrac{{\cal D}\yyy'+{\cal F}_{22} \yyy}{{\cal F}_{21}}
		\,. \label{odefory2}\end{eqnarray}\end{adjustwidth}
As shown in Appendix \ref{appendixsturmstability} these can be written as a Sturm-Liouville problem and solved using a Schwarzian approach. 

There are two essential differences compared to the classical Sturm-Liouville problems, of e.g. Quantum Mechanics, as the examples examined so far.  
The first is that the eigenvalue enters nonlinearly in the problem and in both functions $p$ and $q$. The second is that we work in complex domain, i.e., all known and unknown functions are complex. This is unavoidable if we are interested to find unstable modes in which case the eigenvalue is a complex number. 
(One can follow either temporal, or spatial approach, which correspond to real wavevector and complex frequency, or real frequency and complex wavevector, respectively.)
The methods remain the same though.  

There are many variants of the procedure: working with $\xxx$ and Schwarzian $g$ approach, with $\xxx$ and Schwarzian $\Phi$ approach, 
with $\yyy$ and Schwarzian $g$ approach, with $\yyy$ and Schwarzian $\Phi$ approach.  
All are equivalent to each other. One could think that the Schwarzian $\Phi$ approaches describe more naturally oscillations of the eigenfunctions, but this is not the case. In the complex domain the oscillations do not lead to zeros of the unknown complex function, since this requires its real and imaginary parts to vanish simultaneously. 
On the contrary, $g$ approaches are slightly preferable because the corresponding differential equations are simpler and, more importantly, the quantization condition does not involve a trigonometric function (as a result it does not contain an arbitrary integer).
In real space, as in most examples presented so far, the $g$ approach is problematic because the function $F$ becomes infinite at some points. In the complex domain though, this happens only at poles, which are automatically circumvented during the numerical integration.\endnote{{When we integrate a differential equation in the complex domain we need infinite accuracy to hit a pole, something impossible numerically. For this reason the integration passes without problem close to poles. See a related discussion in Appendix C of \cite{2024Univ...10..183V}.}}  

As explained in \cite{2024Univ...10..183V}, only the ratio $Y=\dfrac{\xxx}{\yyy}$ is involved in the boundary conditions that determine the dispersion relation of a stability problem, not the functions $\xxx$, $\yyy$ separately. The function $Y$ is always continuous, even at interfaces (discontinuities) of the unperturbed state. 
For this reason, if we choose the Sturm-Liouville for $\xxx$ it is advantageous to replace $\xxx'/\xxx$ with the continuous function $Y$, using the second from Equations (\ref{odefory1}).
Similarly, if we choose the Sturm-Liouville for $\yyy$ it is advantageous to replace $\yyy'/\yyy$ with the continuous function $Y$, using the second from Equations (\ref{odefory2}).
Essentially we follow the minimalist approach of Ref. \cite{2024Univ...10..183V} and work with $Y$ alone by integrating the equation 
\begin{eqnarray}
\dfrac{dY}{d\varpi}=\dfrac{{\cal F}_{21}}{{\cal D}}Y^2+\dfrac{{\cal F}_{22}-{\cal F}_{11}}{{\cal D}}Y-\dfrac{{\cal F}_{12}}{{\cal D}}
\,. \label{eqforY} \end{eqnarray}
The Schwarzian approach splits the $Y$ function in two parts and helps in expressing the asymptotic boundary conditions in a convenient way. 

Details and all the needed equations for all variants can be found in Appendix~\ref{appendixsturmstability}. 
The asymptotic behavior near the axis is discussed in Appendix~\ref{appendixasymptaxis}.
According to this, the formulation with $\yyy$ should be avoided in the case $m=0$ when the quantization condition is applied on the axis. All other variants are equivalently applicable.

Here we summarize the procedure for two variants, in order to use them in an example case.

\subsection{Formulation with $\xxx$ and the Schwarzian $\Phi$ Approach}

In this case the ratio $Y$ and the eigenfunctions are given by
\be 
\dfrac{1}{Y}={\cal Y}_4-{\cal Y}_3\cot\dfrac{\Phi_1+C}{2} \,, \quad 
\xxx
\propto\dfrac{1}{\sqrt{{\cal Y}_3 \, e^{\int \frac{{\cal F}_{11}+{\cal F}_{22}}{{\cal D}} d\varpi }}}\sin\dfrac{\Phi_1+C}{2}   
\,, \quad  \yyy=\dfrac{1}{Y}\xxx\,, \label{eqeigenscphiap1} \ee 
where $\Phi_1$, ${\cal Y}_3$ and ${\cal Y}_4$ are a particular solution (with arbitrary initial conditions) of the system of Equations
\begin{adjustwidth}{-3cm}{0cm}\begin{eqnarray} 
	{\cal Y}_4'= -\dfrac{{\cal F}_{21}}{{\cal D}}
	-\dfrac{{\cal F}_{22}-{\cal F}_{11}}{\cal D}{\cal Y}_4 +\left({\cal Y}_4^2-{\cal Y}_3^2\right)\dfrac{{\cal F}_{12}}{{\cal D}}
	\,, \quad
	{\cal Y}_3'= 2{\cal Y}_3 {\cal Y}_4\dfrac{{\cal F}_{12}}{\cal D}+\dfrac{{\cal F}_{11}-{\cal F}_{22}}{\cal D} {\cal Y}_3 
	\,, \quad
	\Phi_1'=2{\cal Y}_3\dfrac{{\cal F}_{12}}{\cal D}\,. 
		\label{eqsscphiap1}
\end{eqnarray}\end{adjustwidth}

The application to each problem with specific boundary conditions is obvious:
\begin{itemize}
	\item
	Suppose we have a jet in $0\le\varpi\le\varpi_j$ with known $Y_{\rm BC}$ at $\varpi_j^+$. 
	We integrate Equations~(\ref{eqsscphiap1}) starting from $\varpi_j$ with arbitrary conditions for $\Phi_1$, ${\cal Y}_3$, ${\cal Y}_4$ and the free additive constant $C$ chosen as $Y|_{\varpi_j}=Y_{\rm BC}$ is satisfied, with $Y|_{\varpi_j}$ given by the first of Equations~(\ref{eqeigenscphiap1}). The eigenvalues are found from the quantization condition on the axis $\sin\dfrac{\Phi_{1 axis}+C}{2}=0$ 
	(the condition ${\cal Y}_3\to 0$ on the axis will automatically satisfied by the solution of the differential equations).
	Note that in case of internal discontinuities of the unperturbed state inside the jet we keep integrating the equations passing each of them keeping $\Phi_1$, ${\cal Y}_3$, ${\cal Y}_4$ continuous.
	\item
	Suppose we want to find numerically the solution for $0\le\varpi\le\infty$. 
	We integrate Equations (\ref{eqsscphiap1}) starting from an arbitrary point with arbitrary conditions for $\Phi_1$, ${\cal Y}_3$, ${\cal Y}_4$, toward both directions, i.e. toward $\varpi=0$ and toward $\varpi=\infty$. The quantization condition $\sin\dfrac{\Phi_{1 \infty}-\Phi_{1 axis}}{2}=0$ gives the eigenvalues.
	If we additionally need to find the eigenfunctions we specify the constant $C$ from either $\sin\dfrac{\Phi_{1 \infty}+C}{2}=0$ or $\sin\dfrac{\Phi_{1 axis}+C}{2}=0$.
\end{itemize}

\subsection{Formulation with $\xxx$ and the Schwarzian $g$ Approach}

In this case the ratio $Y$ and the eigenfunctions are given by
\be 
\dfrac{1}{Y}={\cal Y}_4-\dfrac{e^{-2{\cal Y}_3}}{g_1+C_2/C_1} \,, \quad 
\xxx 
\propto\dfrac{g_1+C_2/C_1}{ e^{-{\cal Y}_3+\int \frac{{\cal F}_{11}+{\cal F}_{22}}{2{\cal D}} d\varpi }} 
\,, \quad  \yyy=\dfrac{1}{Y}\xxx\,, \label{eqeigenscgap1} \ee 
where ${\cal Y}_4$, ${\cal Y}_3$ and $g_1$ are a particular solution (with arbitrary initial conditions) of the Equations   
\begin{adjustwidth}{-2.0cm}{0cm}\begin{eqnarray} 
{\cal Y}_4'=-\dfrac{{\cal F}_{21}}{{\cal D}}-\dfrac{{\cal F}_{22}-{\cal F}_{11}}{\cal D}{\cal Y}_4+\dfrac{{\cal F}_{12}}{\cal D}{\cal Y}_4^2
\,, \quad 
{\cal Y}_3'=-{\cal Y}_4 \dfrac{{\cal F}_{12}}{\cal D}+\dfrac{{\cal F}_{22}-{\cal F}_{11}}{2{\cal D}}  
\,, \quad		
g_1'=\dfrac{{\cal F}_{12}}{\cal D} e^{-2{\cal Y}_3}		
\,. 
\label{eqsscgap1}
\end{eqnarray}\end{adjustwidth}

The application to each problem with specific boundary conditions can be done following the same steps as in the $\Phi$ approach, with the boundary conditions on the axis and at infinity replaced by $g_{1 axis}+C_2/C_1=0$, and $g_{1\infty}+C_2/C_1=0$, respectively.
(If both apply, the eigenvalues can be found by the single condition $g_{1\infty}-g_{1 axis}=0$.)

\section{An Example for the Stability of Astrophysical Jets}\label{secexamplestability}

In this section we apply the Schwarzian approach to the stability of a particular non relativistic jet model considered by Cohn in Ref. \cite{1983ApJ...269..500C}.
The reason behind this choice is mostly connected to the fact that there are analytical solutions for some choices of the parameters and we can use them to test the results of the new method.  

In the unperturbed state, we assume that there is a uniform cylindrical jet of radius $\varpi_j$ consisting of a hydrodynamic fluid with constant polytropic index $\Gamma=5/3$, density $\rho_j$, sound velocity $c_{sj}$, and pressure $P_j=\dfrac{\rho_jc_{sj}^2}{\Gamma}$, moving with bulk velocity $V_j\hat z=Mc_{sj}\hat z$ ($\hat z$ is the symmetry axis and we use cylindrical coordinates $\varpi,\phi,z$).
The environment of the jet is assumed to be a cold, static, ideal, magnetized plasma, with constant density $\rho_e=\eta\rho_j$, and azimuthal magnetic field $\bm B=\dfrac{I}{\varpi}\hat\phi$ with constant $I$.
This field corresponds to a surface current at the interface $\varpi=\varpi_j$ whose value is connected to the internal pressure through the pressure balance $P_j=\dfrac{I^2}{2\varpi_j^2}$. 

We use units for which lengths are measured in $\varpi_j$, wavelengths in $1/\varpi_j$, velocities in $c_{sj}$, frequencies and growth rates in $c_{sj}/\varpi_j$, 
densities in $\rho_j$, and a factor $\sqrt{4\pi}$ is absorbed in the magnetic field (Lorentz–Heaviside units). 
The expressions of the various ${\cal F}_{ij}/{\cal D}$ needed for this non-relativistic ideal magnetohydrodynamic stability problem can be found in Section 5.6 of the Ref. \cite{2023Univ....9..386V} and are summarized in Appendix \ref{appendixnonrelmhdeqs}.

The dimensionless parameters that fully determine the unperturbed state are
the Mach number $M$ in the jet (velocity of the jet in units of the sound velocity in the jet) and the density ratio $\eta$ (density of the environment over the density of the jet).
We assume in the following $M=1$ and $\eta=0.01$. 

\begin{figure}
	\begin{minipage}{0.45\textwidth}
		\includegraphics[width=1\textwidth]{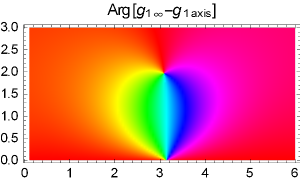} 
		
		\vspace{20pt}
		\includegraphics[width=1\textwidth]{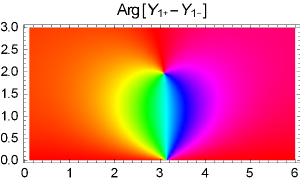}
		
		\vspace{20pt}
		\includegraphics[width=1\textwidth]{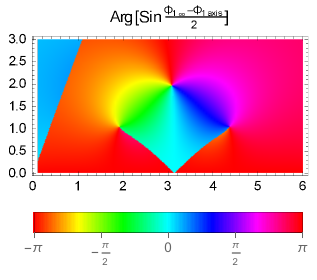}
	\end{minipage}
	\hfill
	\begin{minipage}{0.45\textwidth}
		\includegraphics[width=0.9\textwidth]{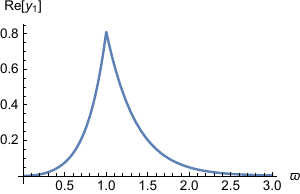} 
		
		\vspace{10pt}
		\includegraphics[width=0.9\textwidth]{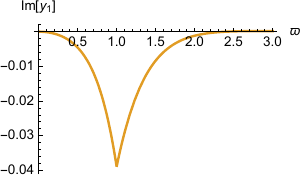}
		
		\vspace{10pt}
		\includegraphics[width=0.9\textwidth]{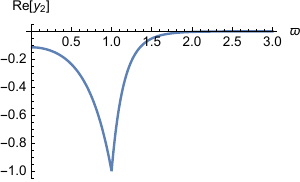} 
		
		\vspace{10pt}
		\includegraphics[width=0.9\textwidth]{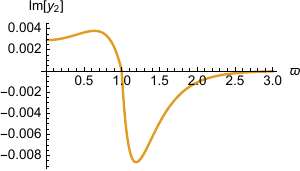}
	\end{minipage}
	\caption{Left column, upper panel: The Spectral Web for the Schwarzian $g$ approach starting the integration from the jet surface with initial conditions 
		$g|_{\varpi=1}=0$, ${\cal Y}_3|_{\varpi=1}=0$, ${\cal Y}_4|_{\varpi=1}=0$.
		Left column, middle panel: The Spectral Web using the analytical expressions of $Y$ inside and outside the jet. 
		Left column, lower panel: The Spectral Web for the Schwarzian $\Phi$ approach starting the integration from the jet surface with initial conditions 
		$\Phi_1|_{\varpi=1}=0$, ${\cal Y}_3|_{\varpi=1}=1$, ${\cal Y}_4|_{\varpi=1}=0$.
		Right column: The eigenfunctions for the eigenvalue $\omega\approx 3.08+i1.97$ with arbitrary normalization (we can freely multiply both $\xxx=\Re\xxx+i\Im\xxx$ and $\yyy=\Re\yyy+i\Im\yyy$ with the same arbitrary complex constant -- their ratio $Y$ is uniquely defined).} 
	\label{figcohnM1m0eta0.01}
\end{figure} 

Suppose we follow the temporal approach and need to find the eigenvalues and eigenfunctions of unstable modes ($\propto e^{i(m\phi+kz-\omega t)} $ with $\Im\omega>0$) corresponding to wavelength $k=\pi$, $m=0$.
Applying the Schwarzian $g$ approach we start the integration of Equations~(\ref{eqsscgap1}) from the jet surface toward smaller and larger $\varpi$. 
The eigenvalues $\omega$ are the roots of the quantization condition $ g_{1 \infty}-g_{1 axis} =0$. 
Following the method presented in Section 3 of Ref. \cite{2024Univ...10..183V}
the roots of this complex function can be found through the isocontours of 
$\Psi={\rm Arg}\left[g_{1 \infty}-g_{1 axis}\right]$ in the $\omega$ plane, the so-called Spectral Web.
This term was introduced by \cite{Goedbook2} for a similar map showing ``solution paths'' and ``conjugate paths'', which correspond to particular isocontours of $\Psi$, and generalized by \cite{2024Univ...10..183V} to the map of $\Psi$ in which the roots can be distinguished from poles. 
The roots are seen as positive line charges in this map, i.e. points of discontinuities around which the $\Psi$ increases as we move counterclockwise. Poles of the complex function are also seen in the map, as negative line charges, and they can be distinguished from the roots because around poles the $\Psi$ decreases as we move counterclockwise.

The left upper panel of Figure \ref{figcohnM1m0eta0.01} shows the resulting Spectral Web. Evidently there is one root $\omega\approx 3.08+i1.97 $. 
(Note that any choice of initial conditions at the starting point of the integration leads to the same result for the roots, i.e., the eigenvalues, although the Spectral Web and the positions of poles -- if they exist -- in general differ.)

The left middle panel of Figure \ref{figcohnM1m0eta0.01} shows the Spectral Web resulting from the
boundary condition $Y|_{\varpi=\varpi_i^+}-Y|_{\varpi=\varpi_i^-}=0$ which is simply the continuity of $Y$ at the jet surface,
using the analytical expressions that exist for this particular model.\endnote{{
The analytical solution for the jet interior $\varpi<\varpi_j$, corresponding to finite eigenfunction on the axis, is $Y=-\dfrac{\lambda\varpi}{\rho_j\omega_0^2}\dfrac{J_1(\lambda\varpi)}{J_0(\lambda\varpi)}$ where $\lambda=\sqrt{\omega_0^2/c_{sa}^2-k^2}$, and $\omega_0=\omega-kV_j$. For the environment $\varpi>\varpi_j$ the solution that represents outgoing wave with decreasing amplitude has
$Y=\dfrac{\varpi^4}{I^2} \left[-\dfrac{{\rm z}^2 U}{(4-{\rm z})U+2{\rm z} dU/d{\rm z}}-1\right]^{-1}$, where ${\rm z}= \dfrac{- i\omega}{|I|/\sqrt{\rho_e}}\varpi^2$ and $U(a,b,{\rm z})$ is Tricomi's (confluent hypergeometric) function satisfying the Kummer equation 
$ {\rm z}\dfrac{d^2U}{d{\rm z}^2}+(b-{\rm z})\dfrac{dU}{d{\rm z}}-aU=0$ with $a=\dfrac{3}{2}+ i\dfrac{k^2|I|/\sqrt{\rho_e}}{4\omega}$ and $b=3$. Besides Ref. \cite{1983ApJ...269..500C}, details for the interior and exterior solution, respectively, can be found in Sections 5.1 and 5.4.2 of \cite{2023Univ....9..386V} (adapting these results to the non relativistic present case).}} 
The resulting eigenvalue is identical to the one found using the Schwarzian approach.  

The eigenfunction can be directly found from Equations (\ref{eqeigenscgap1}), knowing the eigenvalue and choosing $C_2/C_1=-g_{1 axis}$. The result is showing in the right column of Figure \ref{figcohnM1m0eta0.01}.

Note that in both asymptotic regimes, near the axis and at large distances, the function $e^{-{\cal Y}_3}$ vanish and the eigenfunctions approach constant values. Thus, it is sufficient to perform the integration in the region where these functions vary significantly. For the case shown in Figure \ref{figcohnM1m0eta0.01} it was done in the interval $0.01<\varpi<10$.

Applying also the Schwarzian $\Phi$ approach the eigenvalues $\omega$ are the roots of the quantization condition $\sin\dfrac{\Phi_{1 \infty}-\Phi_{1 axis}}{2}=0$. 
The Left bottom panel of Figure \ref{figcohnM1m0eta0.01} shows the corresponding Spectral Web. Apart from the root at $\omega\approx 3.08+i1.97 $
in this case there are two poles and some curves of discontinuities which correspond to crossing of poles during the integration. The existence and the position of the poles and the curves of discontinuity depend on the initial conditions, but the root is always the same.
 
The eigenfunction $Y$ corresponding to a particular eigenvalue is the same no matter which approach we use to find it. The eigenfunction $\xxx$ is also the same apart from a multiplication complex constant which is free due to the linearity of the problem. The same multiplication constant appears in $\yyy$, in agreement with the ratio $Y=\xxx/\yyy$ being uniquely determined.

\begin{figure}
	\includegraphics[width=0.4\textwidth]{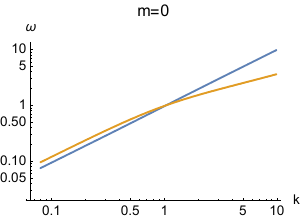} 
	\hfill 
	\includegraphics[width=0.55\textwidth]{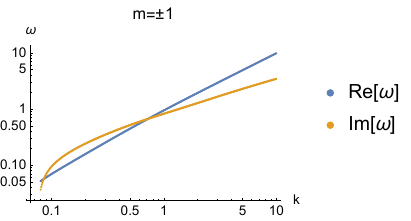}
	\caption{The dispersion relation for the model with $M\equiv \dfrac{V_j}{c_{sj}}=1$, $\eta\equiv\dfrac{\rho_e}{\rho_j}=0.01$, and $m=0$ (left), $m=\pm 1$ (right).} 
	\label{figdispersionallk}
\end{figure}  

Repeating the process for other values of $k$ we find the dispersion relation shown in the left panel of Figure \ref{figdispersionallk}.

Using the Schwarzian approach we can extend the results in the parametric space to values for which no analytical solutions exist. It is beyond the scope of this paper to carry out this interesting task and analyze the physics of the solutions, we only show the dispersion relation for the $m=\pm 1$ non-axisymmetric modes in the right panel of Figure \ref{figdispersionallk} (the result is independent on the sign of $m$ in this particular case).  

\section{Conclusions}\label{secconclusions}

In this work we develop a novel way to solve Sturm-Liouville problems by following the minimalist/Schwarzian approach, particularly useful when the boundaries are at asymptotic distances. The minimalist approach is a first step toward economy and efficiency, since it reduces the differential equation to first order. The second step is the Schwarzian approach. Although the resulting differential equation is nonlinear, the properties of the Schwarzian derivatives allow to split its general solution in two parts. The first part is any particular solution, and the form of the second part can be used to explore under which conditions the sum is non-diverging asymptotically. 

Essentially the Schwarzian approach finds the non-diverging solution from any particular solution, and simultaneously the eigenvalues for which this is possible.
It can be seen as an alternative to other methods to choose the non-diverging solution of a second-order linear differential equation, e.g. the termination of series found using the Frobenius method in quantum mechanics, which however, can be applied to simple cases only.
The new approach can be applied to any case, to stability problems as well, in which the eigenvalue enters nonlinearly in both functions $p$ and $q$. 
The application of the method requires numerical integration of ordinary differential equations; in cases where the boundaries are at asymptotic distances it is sufficient to continue the integration as long as the resulting unknown functions vary significantly. 

Two Schwarzian approaches are presented, the $g$ and $\Phi$, that correspond to different ways to split the general solution in two parts. Although they are mathematically equivalent, they have some practical differences. The $g$ approach is simpler, but cannot trivially handle the infinities in $f'/f$ when the function $f$ is real and oscillates, as in the examples presented in relation with the Schr\"odinger equation in quantum mechanics. 
Nevertheless, as we saw in the example for the quantum Morse potential it can still be used by working in the complex domain, even though the eigenvalues are real.
When $f$ is a complex function, as in stability problems, there is no problem with infinities since zero of a denominator (pole of $f'/f$) requires fine-tuning to make both parts, real and imaginary, to vanish simultaneously. 

The $\Phi$ approach can be seen as a transformation of the $g$ approach, but successfully handles the mentioned infinities in $f'/f$ when the function $f$ is real and oscillates. One could think that the function $\Phi$ is connected with the phase of the oscillations. This is true in some cases, but not always. The form of the corresponding eigenfunction $\propto \dfrac{1}{\sqrt{\F_2}}\sin\dfrac{\Phi+C}{2}$ represents the superposition of two oppositely moving waves, but in general the amplitude is highly variable, so the result is not a standing wave in general. It is an open question if other Schwarzian approaches, corresponding to different transformations and different ways of splitting the general solution to two, can be more closely related to the oscillations of the eigenfunctions, especially in stability problems. Another interesting question is related to the freedom to choose the initial condition that determines the particular solution. Is there an optimal choice? 

Besides questions related to the deeper understanding of the symmetries behind the quantization condition and the somewhat unexpected existence of the method itself, its practical use is obvious. It helps solving difficult problems with relative ease, without needing to work the asymptotic behaviors e.g. near the symmetry axis and near infinity in cylindrical jet problems, and without the need to integrate the equations at very small and very large distances.

\vspace{6pt} 




\funding{This research received no external funding.}

\dataavailability{This research is analytical; no new data were generated or analyzed. If needed, more details on the study and the numerical results will be shared on reasonable request to the author.} 

\acknowledgments{Heartfelt \ref{figcohnM1m0eta0.01} to Christina for silent support.}

\conflictsofinterest{The author declares no conflicts of interest.} 


%


\appendixtitles{yes} 
\appendixstart
\appendix


\section{Properties of the Schwarzian Derivative}\label{appendixschwarzian}

The Schwarzian derivative of a function $g(x)$, in general complex and with complex argument, is defined as 
\be
\{g, x\} \equiv \frac{g'''}{g'} - \frac{3}{2}\left(\frac{g''}{g'}\right)^2 \,,
\ee
where prime denotes derivative with respect to $x$.

An important property of this derivative is that if $\B_1$ and $\B_2$ are two linearly independent solutions of 
\be \B''+\kappa^2 \B=0
\label{eqforbapp}
\ee 
with $\kappa $ a function of $x$, 
then their quotient $g=\dfrac{\B_1}{\B_2}$ satisfies the Schwarz equation 
\be
\{g, x\} = 2\kappa^2
\,. \ee 
Conversely, if $g$ is a solution of $\{g, x\}= 2\kappa^2$ then $\B_1\propto \dfrac{1}{\sqrt{(1/g)'}} $ and $\B_2\propto \dfrac{1}{\sqrt{g'}} $
are two independent solutions of $\B''+\kappa^2\B=0$.
Thus, the general solution of Equation (\ref{eqforbapp}) is
\be 
\B=\dfrac{C_1g+C_2}{\sqrt{g'}}
\,. \ee 
The proof can be done as in Ref. \cite{Hillebook}. 
Direct substitution shows that $\B_{1,2}$ are solutions of Equation~(\ref{eqforbapp}). Also by substitution we find that their Wronskian $\B_1\B_2'-\B_2\B_1'$ is constant, so they are indeed linearly independent. 

In fact every 2nd-order linear differential equation can be written in the form of Equation (\ref{eqoriginalsl}), can be transformed to a ``variable frequency oscillator'' in the form of Equation~(\ref{eqforb}), and its general solution can be written through a particular solution of the equation involving the Schwarzian, as in Equation~(\ref{eqsolforb}).

Some other useful properties follow
($a$, $b$, $c$, $d$ are complex constants): 
\be 
\left\{\dfrac{ax+b}{cx+d}, x\right\} =0
\,,
\ee
\be 
\left\{\dfrac{ag+b}{cg+d}, x\right\} =\{g, x\} 
\ee
(a M\"obius transformation leaves the Schwarzian unchanged),
\be 
\{g , x\}  =\{g, \Phi\}\left(\dfrac{d\Phi}{dx}\right)^2 +\{\Phi, x\}
\label{eqchainrule}
\ee
(can be thought as a chain rule for $\{g\circ \Phi, x\}$),
\be 
\{g, x\}=\left\{g,\dfrac{ax+b}{cx+d}\right\} \, \dfrac{(ad-bc)^2}{(cx+d)^4} 
\ee
(the chain rule for $\Phi=\dfrac{ax+b}{cx+d}$).

\section{Sturm-Liouville Formulation for Stability Problems}\label{appendixsturmstability}

In the next two subsections of this Appendix we analyze the Sturm-Liouville formulation for the functions $\xxx$ and $\yyy$ separately. These are though as functions of the cylindrical radius $\varpi$ and primes denote derivatives with respect to that variable.   

\subsection{Formulation with $\xxx$}

Equations 
(\ref{odefory1})
are equivalent to a  Sturm-Liouville problem for $\xxx$
\be
\xxx''+2\gamma_1\xxx'+b_1\xxx=0
\Leftrightarrow 
\left(e^{\int 2\gamma_1 d\varpi } \xxx'\right)'+e^{\int 2\gamma_1 d\varpi } b_1 \xxx=0\,, 
\ee
\be 
\gamma_1= \dfrac{{\cal F}_{11}+{\cal F}_{22}}{2{\cal D}}+\dfrac{{\cal F}_{12}}{2{\cal D}}\left(\dfrac{{\cal D}}{{\cal F}_{12}}\right)' \,, \quad 
b_1= \dfrac{{\cal F}_{11} {\cal F}_{22}-{\cal F}_{12} {\cal F}_{21}}{{\cal D}^2}
+\dfrac{{\cal F}_{12}}{{\cal D}}\left(\dfrac{{\cal F}_{11}}{{\cal F}_{12}}\right)' \,, 
\label{eqgamma1kappa1}\ee
and the expression for $\yyy$ (as function of $\xxx$)
\be
\yyy=-\dfrac{{\cal D} }{{\cal F}_{12}}\xxx'-\dfrac{ {\cal F}_{11}}{{\cal F}_{12}} \xxx \,.
\ee 
The corresponding ``variable frequency oscillator'' is
\be \left(e^{\int \gamma_1 d\varpi }\xxx \right)''+\kappa_1^2\left(e^{\int \gamma_1 d\varpi }\xxx \right)=0 \,, \quad \kappa_1^2=b_1-\gamma_1^2-\gamma_1' \,.
\ee 
It's solution in the Schwarzian $g$ approach is
\be \xxx
=\dfrac{C_1g_1+C_2}{\sqrt{g_1'\dfrac{{\cal D}}{{\cal F}_{12}}e^{\int \frac{{\cal F}_{11}+{\cal F}_{22}}{{\cal D}} d\varpi }}} 
\,, \quad 
\{g_1, \varpi \} = \frac{g_1'''}{g_1'} - \frac{3}{2}\left(\frac{g_1''}{g_1'}\right)^2=2\kappa_1^2
\,. \label{eqvarfrxxx}
\ee
Writing $\dfrac{1}{Y}=\dfrac{\yyy}{\xxx}=-\dfrac{{\cal D} }{{\cal F}_{12}}\dfrac{\xxx'}{\xxx}-\dfrac{ {\cal F}_{11}}{{\cal F}_{12}} $ as 
\be \dfrac{1}{Y}={\cal Y}_4-\dfrac{e^{-2{\cal Y}_3}}{g_1+C_2/C_1} 
\ee 
we end-up with the following system
\begin{adjustwidth}{-\extralength}{0cm}\begin{eqnarray} 
	{\cal Y}_4'=-\dfrac{{\cal F}_{21}}{{\cal D}}-\dfrac{{\cal F}_{22}-{\cal F}_{11}}{\cal D}{\cal Y}_4+\dfrac{{\cal F}_{12}}{\cal D}{\cal Y}_4^2\,,
	\quad {\cal Y}_3'= -{\cal Y}_4 \dfrac{{\cal F}_{12}}{\cal D}+\dfrac{{\cal F}_{22}-{\cal F}_{11}}{2{\cal D}} \,, 
	\quad
	g_1'=\dfrac{{\cal F}_{12}}{\cal D} e^{-2{\cal Y}_3} 
	 \,. 
\end{eqnarray}\end{adjustwidth}
Similarly in the Schwarzian $\Phi$ approach
\be \xxx
\propto\dfrac{1}{\sqrt{\Phi_1'\dfrac{{\cal D}}{{\cal F}_{12}}e^{\int \frac{{\cal F}_{11}+{\cal F}_{22}}{{\cal D}} d\varpi }}}\sin\dfrac{\Phi_1+C}{2}  
\,, \quad   \dfrac{\Phi_1'''}{\Phi_1'}-\dfrac{3\Phi_1''^2}{2\Phi_1'^2}+\dfrac{\Phi_1'^2}{2} =2\kappa_1^2
\,, \label{eqvarfrxxxphi}\ee 
and by writing $\dfrac{1}{Y}=\dfrac{\yyy}{\xxx}=-\dfrac{{\cal D} }{{\cal F}_{12}}\dfrac{\xxx'}{\xxx}-\dfrac{ {\cal F}_{11}}{{\cal F}_{12}} $ as 
\be \dfrac{1}{Y}={\cal Y}_4 -{\cal Y}_3\cot\dfrac{\Phi_1+C}{2}
\label{eqY34}
\ee 
we end-up with the following system
\begin{adjustwidth}{-\extralength}{0cm}\begin{eqnarray} 
{\cal Y}_4'= -\dfrac{{\cal F}_{21}}{{\cal D}}
-\dfrac{{\cal F}_{22}-{\cal F}_{11}}{\cal D}{\cal Y}_4 +\left({\cal Y}_4^2-{\cal Y}_3^2\right)\dfrac{{\cal F}_{12}}{{\cal D}}  \,,
\quad
{\cal Y}_3'= 2{\cal Y}_3 {\cal Y}_4\dfrac{{\cal F}_{12}}{\cal D}+\dfrac{{\cal F}_{11}-{\cal F}_{22}}{\cal D} {\cal Y}_3 
\,,\quad
\Phi_1'=2{\cal Y}_3\dfrac{{\cal F}_{12}}{\cal D} 
\,. 
\label{eqsscphiap}
\end{eqnarray}\end{adjustwidth}

\subsection{Formulation with $\yyy$}

Equations 
(\ref{odefory2})
are equivalent to a  Sturm-Liouville problem for $\yyy$
\be
\yyy''+2\gamma_2\yyy'+b_2\yyy=0
\Leftrightarrow 
\left(e^{\int 2\gamma_2 d\varpi } \yyy'\right)'+e^{\int 2\gamma_2 d\varpi } b_2 \yyy=0\,, 
\ee
\be 
\gamma_2= \dfrac{{\cal F}_{11}+{\cal F}_{22}}{2{\cal D}}+\dfrac{{\cal F}_{21}}{2{\cal D}}\left(\dfrac{{\cal D}}{{\cal F}_{21}}\right)' \,, \quad 
b_2= \dfrac{{\cal F}_{11} {\cal F}_{22}-{\cal F}_{12} {\cal F}_{21}}{{\cal D}^2}
+\dfrac{{\cal F}_{21}}{{\cal D}}\left(\dfrac{{\cal F}_{22}}{{\cal F}_{21}}\right)' \,, 
\label{eqgamma2kappa2} \ee
and the expression for $\xxx$ (as function of $\yyy$)
\be
\xxx=-\dfrac{{\cal D} }{{\cal F}_{21}}\yyy'-\dfrac{ {\cal F}_{22}}{{\cal F}_{21}} \yyy \,.
\ee 
The corresponding ``variable frequency oscillator'' is
\be \left(e^{\int \gamma_2 d\varpi }\yyy \right)''+\kappa_2^2\left(e^{\int \gamma_2 d\varpi }\yyy \right)=0 \,, \quad \kappa_2^2=b_2-\gamma_2^2-\gamma_2' \,.
\ee 
It's solution in the Schwarzian $g$ approach is
\be \yyy
=\dfrac{C_1g_2+C_2}{\sqrt{g_2'\dfrac{{\cal D}}{{\cal F}_{21}}e^{\int \frac{{\cal F}_{11}+{\cal F}_{22}}{ {\cal D}} d\varpi }}} 
\,, \quad 
\{g_2, \varpi \} = \frac{g_2'''}{g_2'} - \frac{3}{2}\left(\frac{g_2''}{g_2'}\right)^2=2\kappa_2^2
\,. \label{eqvarfryyy}
\ee
Writing $Y=\dfrac{\xxx}{\yyy}=-\dfrac{{\cal D} }{{\cal F}_{21}}\dfrac{\yyy'}{\yyy}-\dfrac{ {\cal F}_{22}}{{\cal F}_{21}} $ as 
\be Y={\cal Y}_2 -\dfrac{e^{-2{\cal Y}_1}}{g_2+C_2/C_1}
\ee 
we end-up with the following system
\begin{adjustwidth}{-\extralength}{0cm}\begin{eqnarray} 
{\cal Y}_2'= \dfrac{{\cal F}_{21}}{\cal D}{\cal Y}_2^2+\dfrac{{\cal F}_{22}-{\cal F}_{11}}{\cal D}{\cal Y}_2- \dfrac{{\cal F}_{12}}{{\cal D}} 		 
\,, \quad
{\cal Y}_1'= -{\cal Y}_2\dfrac{{\cal F}_{21}}{\cal D}-\dfrac{{\cal F}_{22}-{\cal F}_{11}}{2{\cal D}}  
\,, \quad
g_2'=\dfrac{{\cal F}_{21}}{\cal D} e^{-2{\cal Y}_1}
\,. 
\end{eqnarray}\end{adjustwidth}
Similarly in the Schwarzian $\Phi$ approach
\be \yyy
\propto \dfrac{1}{\sqrt{\Phi_2'\dfrac{{\cal D}}{{\cal F}_{21}}e^{\int \frac{{\cal F}_{11}+{\cal F}_{22}}{ {\cal D}} d\varpi }}}\sin\dfrac{\Phi_2+C}{2}  \,, \quad   \dfrac{\Phi_2'''}{\Phi_2'}-\dfrac{3\Phi_2''^2}{2\Phi_2'^2}+\dfrac{\Phi_2'^2}{2} =2\kappa_2^2
\,, \label{eqvarfryyyphi}\ee 
and by writing $Y=\dfrac{\xxx}{\yyy}=-\dfrac{{\cal D} }{{\cal F}_{21}}\dfrac{\yyy'}{\yyy}-\dfrac{ {\cal F}_{22}}{{\cal F}_{21}} $ as 
\be Y={\cal Y}_2-{\cal Y}_1\cot\dfrac{\Phi_2+C}{2}
\label{eqY12}
\ee 
we end-up with the following system
\begin{adjustwidth}{-\extralength}{0cm}\begin{eqnarray} 
{\cal Y}_2'=\left({\cal Y}_2^2-{\cal Y}_1^2\right)\dfrac{{\cal F}_{21}}{{\cal D}} +\dfrac{{\cal F}_{22}-{\cal F}_{11}}{\cal D}{\cal Y}_2- \dfrac{{\cal F}_{12}}{{\cal D}}
\,,
\quad  
{\cal Y}_1'= 2{\cal Y}_1{\cal Y}_2\dfrac{{\cal F}_{21}}{\cal D}+\dfrac{{\cal F}_{22}-{\cal F}_{11}}{\cal D} {\cal Y}_1 \,, 
\quad \Phi_2'=2{\cal Y}_1\dfrac{{\cal F}_{21}}{\cal D} 
 \,. 
\label{systemy2scwphi}
\end{eqnarray}\end{adjustwidth}

The formulations with $\xxx$ and $\yyy$ are of course equivalent. Actually by replacing ${\cal Y}_1=-{\cal Y}\sin\dfrac{\Phi_1-\Phi_2}{2}$, ${\cal Y}_2={\cal Y}\cos\dfrac{\Phi_1-\Phi_2}{2}$ in Equation~(\ref{eqY12}) we find Equation~(\ref{eqY34}), with ${\cal Y}_3=\dfrac{1}{{\cal Y}}\sin\dfrac{\Phi_1-\Phi_2}{2}$, ${\cal Y}_4=\dfrac{1}{{\cal Y}}\cos\dfrac{\Phi_1-\Phi_2}{2}$ (this corresponds to a M\"obius transformation that leaves the Schwarzian unaffected).

So both formulations are equivalent with the following symmetrical expressions:

\begin{eqnarray}
	Y={\cal Y}\dfrac{\sin\dfrac{\Phi_1+C}{2}}{\sin\dfrac{\Phi_2+C}{2}} \,, \quad
	\yyy= \dfrac{D\sin\dfrac{\Phi_2+C}{2}}{\sqrt{e^{\int \frac{{\cal F}_{11}+{\cal F}_{22}}{ {\cal D}} d\varpi }{\cal Y}\sin\dfrac{\Phi_1-\Phi_2}{2}}}
	\,, \quad \xxx=Y\yyy\,,
	\\
	\Phi_1'=2 \dfrac{{\cal F}_{12}}{\cal D}\dfrac{1}{\cal Y} \sin\dfrac{\Phi_1-\Phi_2}{2}
	\,,
	\quad
	\Phi_2'=-2 \dfrac{{\cal F}_{21}}{\cal D} {\cal Y}\sin\dfrac{\Phi_1-\Phi_2}{2} \,,
	\\
	{\cal Y}'= \left(\dfrac{{\cal F}_{21}}{\cal D}{\cal Y}^2-\dfrac{{\cal F}_{12}}{\cal D} \right)\cos\dfrac{\Phi_1-\Phi_2}{2}+\dfrac{{\cal F}_{22}-{\cal F}_{11}}{\cal D} {\cal Y}
	\,.
\end{eqnarray} 
Asymptotically $\sin\dfrac{\Phi_1-\Phi_2}{2}\to 0 $ and the quantization condition is $\sin\dfrac{\Phi_1+C}{2}\to 0$ (the seemingly equivalent $\sin\dfrac{\Phi_2+C}{2}\to 0$ is not accurate in cases with $m=0$ when the quantization condition is applied on the axis, as discussed in Appendix~\ref{appendixasymptaxis}).

\section{The Stability Problem near the Symmetry Axis $\varpi=0$}\label{appendixasymptaxis}

\begin{itemize}
	\item For $m\neq 0$ near the axis all limits $d_{ij}=\displaystyle\lim_{\varpi\to 0}\dfrac{\varpi {\cal F}_{ij}}{\cal D}$ are constants (given in Appendix B of \cite{2023Univ....9..386V}) and the relations $d_{22}=-d_{11} $, $d_{11}^2+d_{12}d_{21} =m^2$ hold.
	\\
	Using these we can find approximate expressions for the functions given by Equations~(\ref{eqgamma1kappa1}) 
	$\varpi\gamma_1\approx \dfrac{1}{2 }$, $\varpi^2 b_1\approx -m^2$, $\varpi^2 \kappa_1^2\approx \dfrac{1}{4}-m^2$. The solutions of Equations~(\ref{eqvarfrxxx}), (\ref{eqvarfrxxxphi}) behave as 
	$g_1'\propto \varpi^{2|m|-1} $, $\Phi_1'\propto \varpi^{2|m|-1}$, approaching zero as $\varpi\to 0$.
	The approximate solution of Equations~(\ref{eqsscphiap}) is 
	$\Phi_1\approx \Phi_{1 axis}+C_0\varpi^{2|m|}$, ${\cal Y}_3\approx \dfrac{C_0|m|}{d_{12}}\varpi^{2|m|}$, ${\cal Y}_4\approx \dfrac{|m|-d_{11}}{d_{12}}$, and thus
	$\dfrac{1}{Y}\approx -\dfrac{C_0|m|}{2d_{12}}\varpi^{2|m|}\cot\dfrac{\Phi_{1 axis}+C+C_0\varpi^{2|m|}}{2}+\dfrac{|m|-d_{11}}{d_{12}}$.
	For all values of $\Phi_{1 axis}+C$ that do not satisfy the quantization condition 
	the result is the unphysical branch $\dfrac{1}{Y}\approx \dfrac{|m|-d_{11}}{d_{12}}$,
	For $\Phi_{1 axis}+C=2n\pi$ though, the resulting $\dfrac{1}{Y}\approx -\dfrac{|m|+d_{11}}{d_{12}}$ is the acceptable solution (see Equation (5) in \cite{2024Univ...10..183V}).
	\\ Similarly we can find approximate expressions for the functions given by Equations~(\ref{eqgamma2kappa2}) 
	$\varpi\gamma_2\approx \dfrac{1}{2 }$, $\varpi^2 b_2\approx -m^2$, $\varpi^2 \kappa_2^2\approx \dfrac{1}{4}-m^2$. The solutions of Equations~(\ref{eqvarfryyy}), (\ref{eqvarfryyyphi}) behave as
	$g_2'\propto \varpi^{2|m|-1} $, $\Phi_2'\propto \varpi^{2|m|-1}$, approaching zero as $\varpi\to 0$.
\item For $m= 0$ near the axis the constant limits are 
$b_{11}=\displaystyle\lim_{\varpi\to 0}\dfrac{{\cal F}_{11}}{\varpi {\cal D}}$,
$b_{12}=\displaystyle\lim_{\varpi\to 0}\dfrac{{\cal F}_{12}}{\varpi {\cal D}}$,
$b_{21}=\displaystyle\lim_{\varpi\to 0}\dfrac{\varpi {\cal F}_{21}}{{\cal D}}$,
$b_{22}=\displaystyle\lim_{\varpi\to 0}\dfrac{{\cal F}_{22}}{\varpi {\cal D}}$
(see Appendix B of \cite{2023Univ....9..386V}).
\\
Using these we can find approximate expressions for the functions given by Equations~(\ref{eqgamma1kappa1}) 
$\varpi\gamma_1\approx -\dfrac{1}{2 }$, $b_1\approx -b_{12}b_{21}$, $\varpi^2 \kappa_1^2\approx -\dfrac{3}{4}
$, and for the solutions of Equations~(\ref{eqvarfrxxx}), (\ref{eqvarfrxxxphi}) 
$g_1'\propto \varpi $, $\Phi_1'\propto \varpi$ 
(approaching zero as $\varpi\to 0$). 
The approximate solution of Equations~(\ref{eqsscphiap}) is 
$\Phi_1\approx \Phi_{1 axis}+C_0\varpi^2$, ${\cal Y}_3\approx \dfrac{C_0 }{b_{12}}$, ${\cal Y}_4\approx -b_{21}\ln\varpi$, and 
for $\Phi_{1 axis}+C_0=2n\pi$ the resulting $\dfrac{1}{Y}\approx -\dfrac{2}{b_{12}\varpi^2}$ is the acceptable solution (see Equation (6) in \cite{2024Univ...10..183V}).
%
%
%
\\ Similarly we can find approximate expressions for the functions given by Equations~(\ref{eqgamma2kappa2}) 
$\varpi\gamma_2\approx \dfrac{1}{2 }$, $b_2\approx -b_{12}b_{21}+2b_{22}$, $\varpi^2 \kappa_2^2\approx \dfrac{1}{4}$. However in this case the solution of Equation~(\ref{eqvarfryyy}) 
$g_2'\propto \varpi^{-1} $ does not approach zero as $\varpi\to 0$. 
So this method should be avoided when using the quantization condition on the axis.

\end{itemize}

\section{The Stability Problem for Nonrelativistic Cylindrical Jets}\label{appendixnonrelmhdeqs}

If the unperturbed sate has density $\rho_0(\varpi)$, pressure $P_0(\varpi)$, bulk velocity $V_0(\varpi)\hat z$, and magnetic field $\bm B_0=B_{0z}(\varpi)\hat z+B_{0\phi}(\varpi)\hat \phi$  
satisfying the equilibrium condition 
\begin{eqnarray}
\dfrac{d P_0 }{d\varpi}
+ \dfrac{d }{d\varpi}\left(\dfrac{B_{0z}^2}{2}\right)
+ \dfrac{1}{\varpi^2}\dfrac{d }{d\varpi}\left(\dfrac{\varpi^2B_{0\phi}^2}{2}\right)
=0 \,,
\end{eqnarray}  
the linearization of the ideal magnetohydrodynamic equations lead to the system (\ref{systemodes}) with  
\begin{adjustwidth}{-\extralength}{0cm}\begin{eqnarray}  
	-\varpi\dfrac{{\cal F}_{11}}{\cal D} = 
	\dfrac{B_{0\phi}^2 \tilde\kappa^2 
		+2B_{0\phi} k \left(B_{0\phi} k -B_{0z} m/\varpi\right)}{\rho_0 \omega_{\rm co}^2-\left(\bm k_{\rm co} \cdot \bm B_0 \right)^2} \,,
	\quad  \dfrac{{\cal F}_{22}}{\cal D}=-\dfrac{{\cal F}_{11}}{\cal D} \,, 
	\\
	\varpi \dfrac{{\cal F}_{12}}{\cal D} = 
	\dfrac{\tilde\kappa^2 \varpi^2}{\rho_0 \omega_{\rm co}^2-\left(\bm k_{\rm co} \cdot \bm B_0 \right)^2} \,, 
	\quad
	-\varpi\dfrac{{\cal F}_{21}}{\cal D} =
	\rho_0 \omega_{\rm co}^2-\left(\bm k_{\rm co} \cdot \bm B_0 \right)^2 +\dfrac{ B_{0\phi}^2 }{\varpi^2} 
	\dfrac{B_{0\phi}^2\tilde\kappa^2 -4 B_{0z} k \left(\bm k_{\rm co} \cdot \bm B_0 \right)}{\rho_0 \omega_{\rm co}^2-\left(\bm k_{\rm co} \cdot \bm B_0 \right)^2}  
	  \,,
\\
	\tilde\kappa^2 = \dfrac{ \rho_0 \omega_{\rm co}^4} { \left( \rho_0 c_s^2+B_0^2\right) \omega_{\rm co}^2 -c_s^2 \left(\bm k_{\rm co} \cdot \bm B_0\right)^2 }-\bm k_{\rm co}^2
	\,, \quad c_s=\sqrt{\dfrac{\Gamma P_0}{\rho_0}}
	\,, \quad 
	\bm k_{\rm co} = k\hat z
	+\dfrac{m}{\varpi} \hat \phi
	\,, \quad 
	\omega_{\rm co}= \omega-k V_0  
	\,.
\end{eqnarray}\end{adjustwidth}

\begin{adjustwidth}{-\extralength}{0cm}
\printendnotes[custom] 

\reftitle{References}

\vspace{-1mm}
\PublishersNote{}
\end{adjustwidth}

\begin{thebibliography}{999}
	
	\bibitem[Werner O.~Amrein(2005)]{amreinbook}
	Werner O.~Amrein, Andreas M.~Hinz, D.P.P., Ed.
	\newblock {\em Sturm-Liouville Theory: Past and Present}; Birkh\"auser,  2005.
	
	\bibitem[Kravchenko(2020)]{kravchenko2020direct}
	Kravchenko, V.V.
	\newblock {\em Direct and Inverse Sturm-Liouville Problems}; Springer,  2020.
	
	\bibitem[{Goedbloed} et~al.(2019){Goedbloed}, {Keppens}, and
	{Poedts}]{Goedbook2}
	{Goedbloed}, H.; {Keppens}, R.; {Poedts}, S.
	\newblock {\em {Magnetohydrodynamics of Laboratory and Astrophysical Plasmas}};
	Cambridge Univ. Press, Cambridge,  2019.
	
	\bibitem[{Vlahakis}(2023)]{2023Univ....9..386V}
	{Vlahakis}, N.
	\newblock {Linear Stability Analysis of Relativistic Magnetized Jets:
		Methodology}.
	\newblock {\em Universe} {\bf 2023}, {\em 9},~386.
	\newblock {\url{https://doi.org/10.3390/universe9090386}}.
	
	\bibitem[{Vlahakis}(2024)]{2024Univ...10..183V}
	{Vlahakis}, N.
	\newblock {Linear Stability Analysis of Relativistic Magnetized Jets: The
		Minimalist Approach}.
	\newblock {\em Universe} {\bf 2024}, {\em 10},~183.
	\newblock {\url{https://doi.org/10.3390/universe10040183}}.
	
	\bibitem[{Dahl} and {Springborg}(1988)]{1988JChPh..88.4535D}
	{Dahl}, J.P.; {Springborg}, M.
	\newblock {The Morse oscillator in position space, momentum space, and phase
		space}.
	\newblock {\em \jcp} {\bf 1988}, {\em 88},~4535.
	\newblock {\url{https://doi.org/10.1063/1.453761}}.
	
	\bibitem[{Cohn}(1983)]{1983ApJ...269..500C}
	{Cohn}, H.
	\newblock {The stability of a magnetically confined radio jet}.
	\newblock {\em \apj} {\bf 1983}, {\em 269},~500.
	\newblock {\url{https://doi.org/10.1086/161059}}.
	
	\bibitem[{Hille}(1976)]{Hillebook}
	{Hille}, E.
	\newblock {\em {Ordinary differential equations in the complex domain}}; John
	Wiley \& Sons, New York,  1976.
	
\end{thebibliography}
\end{document}